\documentclass[journal]{IEEEtran}

\usepackage{graphicx}
\usepackage{amssymb}
\usepackage{multirow}
\usepackage{indentfirst}
\usepackage{inputenc}
\usepackage{array}
\usepackage{multicol}
\usepackage[switch,mathlines]{lineno}
\usepackage{soul}
\usepackage{graphicx}
\usepackage{cite}
\usepackage[cmex10]{amsmath}
\usepackage{algorithmic}
\usepackage{array}
\usepackage{CJK}
\usepackage[cmex10]{amsmath}
\usepackage{algorithmic}
\usepackage{array}
\usepackage{fixltx2e}
\usepackage{bm}
\usepackage{amssymb}
\ifCLASSOPTIONcompsoc
 \usepackage[caption=false,font=normalsize,labelfont=sf,textfont=sf]{subfig}
\else
 \usepackage[caption=false,font=footnotesize]{subfig}
\fi
\usepackage{url}
\begin{document}

\title{Nonlinear Virtual Inertia Control of WTGs for Enhancing Primary Frequency Response and Suppressing Drive-Train Torsional Oscillations}

\author{{Bi~Liu, Qi~Huang, \IEEEmembership{Senior Member,~IEEE},
    Junbo~Zhao, \IEEEmembership{Senior Member,~IEEE}, Federico~Milano,
    \IEEEmembership{Fellow,~IEEE}, Yingchen~Zhang \IEEEmembership{Senior Member,~IEEE} }
  \thanks{This research work is supported by The National Key Research
    and Development Program of China (No. 2017YFB0902002).
    \par B. Liu and Q. Huang are with the Sichuan Provincial Key Lab of Power System Wide-area Measurement and Control, University of Electronic
    Science and Technology of China, Chengdu, Sichuan, 611731, PR China. (e-mail: liubi\_2006@126.com; whu@uestc.edu.cn).
    \par J. Zhao is with the Department of Electrical and Computer Engineering, Mississippi State University, Starkville, MS 39762, (junbo@ece.msstate.edu).
    \par F. Milano is with the School of Electrical and Electronic
    Engineering, University College Dublin, Dublin, Ireland (e-mail:
    federico.milano@ucd.ie).
    \par Y. Zhang is with National Renewable Energy Laboratory, Golden, CO 80401 USA (e-mail: Yingchen.Zhang@nrel.gov).}}

\markboth{IEEE TRANSACTIONS ON POWER SYSTEMS, 2020}%
{Shell \MakeLowercase{\textit{et al.}}: Bare Demo of IEEEtran.cls for Journals}
\maketitle

\begin{abstract}
Virtual inertia controllers (VICs) for wind turbine generators (WTGs) have been recently developed to compensate the reduction of inertia in power systems. However, VICs can induce low frequency torsional oscillations of the drive train of WTGs.  This paper addresses this issue and develops a new nonlinear VIC based on objective holographic feedbacks theory.  This approach allows transforming the original nonlinear control into a completely controllable system of Brunovsky's type.  Simulation results under various scenarios demonstrate that the proposed method outperforms existing VICs in terms of suppression of low frequency drive-train torsional oscillations, enhancement of system frequency nadir as well as fast and smooth recovery of WTG rotor speed to the original MPP point before the disturbance. The proposed method is also able to coordinate multiple WTGs.
\end{abstract}

\IEEEpeerreviewmaketitle

\section{Introduction}

\subsection{Motivation}  
A relevant challenge associated with the increasing penetration of wind generation is that wind turbine generators (WTGs) do not contribute to the frequency support if no specific control is designed for that \cite{Zhao2017Data, pscc2018}. How to design such a control, however, is not a trivial task. This paper focuses on Virtual inertia control (VIC), which has been developed to emulate the inertia from WTGs for frequency support during an event \cite{Wang2017Control}. Conventional VIC may induce low frequency drive-train torsional oscillations of the wind turbine, thus resulting in a reduction of the service life of WTGs, economic losses and system security issues.  Moreover, there are still unsolved issues of preventing the excessive reduction of the WTG rotor speed at the low wind speed as well as the smooth and fast recovery of WTG rotor speed after the frequency support service.  This paper aims to develop a nonlinear VIC that can enhance primary frequency response, damp out the WTG low frequency drive-train torsional oscillations and achieve fast and smooth WTG rotor speed recovery.

\vspace{-0.3cm}
\subsection{Literature Review}
In the literature, VICs of Type 1 and Type 2 WTGs are developed to provide inertia response for frequency support \cite{Muljadi2012Understanding}. However, there is a lack of flexibility in capturing the maximum wind energy of these two types of WTGs. Today, the most widely used WTGs are Type 3, namely the doubly-fed induction generator (DFIG), and Type 4, namely the fully rated converter-based WTG. The VIC strategies for these converter-iterfaced WTGs can be classified into two main categories: power reserve-based VIC \cite{Vidyanandan2013Primary, Hua2016Analytical} and maximum power point (MPP) tracing-based VIC \cite{Morren2006Inertial, Morren2006Wind, Margaris2012Frequency, Min2016Dynamic, Ullah2008Temporary, Itani2011Short}. The power reserve-based VIC suggests WTG operate at sub-optimal rotor speed via the over speed deloading strategy and it can provide long term frequency support by means of releasing the reserved power. As a side effect, it causes financial loss for wind farm owners. By contrast, the MPP tracing-based VIC releases the kinetic energy stored in the WTG rotor to provide power support during the early stage of frequency contingency. 

The MPP tracing-based VIC has two kinds of control strategies: the predetermined kinetic energy-based VIC \cite{Ullah2008Temporary, Itani2011Short} and frequency dependent VIC \cite{Morren2006Inertial, Morren2006Wind, Margaris2012Frequency, Min2016Dynamic}. The former one releases predetermined WTG rotor kinetic energy after contingency, but this may lead to the secondary frequency dip while restoring the WTG rotor speed. The frequency dependent VIC increases the WTG output active power for frequency support considering the frequency deviation and the rate of frequency deviation. This control, which has constant parameters, may cause the failure of WTG for frequency support under low wind speed and is also subject to the secondary frequency dip.  The popular MPP tracing-based VIC is a frequency dependent VIC with the capability of tuning the parameters \cite{Morren2006Wind, Kayikci2009Dynamic}. 

The VIC of DFIGs requires instant power increment for frequency response \cite{Kang2016Stable}. It may lead to negligible benefit of improving the frequency nadir and cause a low frequency drive-train torsional oscillation of WTG. Since VIC is in fact a function of the frequency deviation and the rate of frequency deviation, the parameters in the function are suggested to be tuned to achieve higher frequency nadir \cite{Vyver2016Droop}. Again, the rotor speed recovery of WTG after frequency support is still subject to the secondary grid frequency dip and low frequency drive-train torsional oscillation of WTG. To deal with that, a time-varying droop characteristics of VIC is proposed for the smooth switch between the frequency support mode and rotor speed recovery mode \cite{Garmroodi2017Frequency}. 

During the rotor speed acceleration duration, the frequency support power of WTG required by VIC decreases not only with time but also with rotor speed \cite{Yang2018Temporary}. This VIC further helps the rapid rotor speed recovery and has the ability of avoiding secondary frequency dip. However, the low frequency drive-train torsional oscillation of WTG at the initial moment of frequency support has not been suppressed. Beside the aforementioned issues, the automatic regulation of frequency support power based on the available kinetic energy of WTG does not fully consider the variations of wind speed and operating rotor speeds. A properly coordinated distribution of the frequency support power among multiple WTGs has not been addressed in existing VICs.

\vspace{-0.3cm}
\subsection{Contributions}
This paper develops a nonlinear VIC to deal with the aforementioned challenges. Specific contributions are as follows.

\begin{itemize}
\item A nonlinear VIC formulation is derived based on the objective holographic feedbacks theory (OHFT) \cite{Hui2008Nonlinear}. The proposed design improves frequency nadir and suppresses low
frequency drive-train torsional oscillation of WTG.  The key idea is to construct the completely controllable Brunovsky system so that the control objectives can be fully satisfied.

\item Modified time-varying and rotor speed dependent functions are developed to automatically regulate the frequency support power of WTGs considering the variations of wind speed and operating rotor speeds. This facilitates smooth and fast recovery of WTG rotor speed to the original MPP point before the disturbance and prevents the secondary frequency dip caused by the sharp switch between frequency support mode and rotor speed recovery mode.

\item The proposed VIC delivers the solution of coordinating the distribution of frequency support power among multiple WTGs, leading to the enhancement of frequency nadir.
\end{itemize}

\vspace{-0.3cm}
\subsection{Organization}
The remainder of this paper is organized as follows. Section II introduces the OHFT and existing VIC control strategies and their related issues.  The proposed nonlinear VIC is also presented in Section II. Extensive simulations are carried out and analyzed in Section III. Section IV concludes the paper.

\vspace{-0.3cm}
\section{OHFT-based Nonlinear VIC of WTGs}
This section outlines the OHFT, presents the proposed nonlinear VIC control for single WTG, and finally, discusses how to extend such a control to the scenario of multiple WTGs.

\subsection{Objective Holographic Feedbacks Theory}
A single-input-multiple-outputs nonlinear system can be described by the follow equations:

\begin{equation}
  \label{Eq.1}
  \begin{aligned}
     \frac{d \bm{x}}{dt} &= \bm{r}(\bm{x})+\bm{s}(\bm{x})u \, , \\
     \bm{y} &= \bm{h}(\bm{x})  \, , 
   \end{aligned}
\end{equation}
where $\bm{x}$ and $\bm{y}$ are the state
and output vectors, respectively; $u$ is the control input, $\bm{r}(\bm{x})$, $\bm{s}(\bm{x})$ and $\bm{h}(\bm{x})$ are vectors of nonlinear functions. If $[y^{*}_{1},y^{*}_{2},...,y^{*}_{m}]$ are the target output variables, define the following multiple-objective equations
\begin{equation}
\label{Eq.2}
  I_i=y_i-y^{*}_{i} \qquad \quad i=1,2,\dots,m \, ,
\end{equation}
where $m$ is the number of output variables and $I_i$ represents
the tracking error of $y_i$. The following conditions have to be satistied:
\begin{equation}
  \lim_{t\rightarrow 0}
  \left|I_i \right|=0   \qquad \quad i=1, 2, \dots, m \, .
  \label{Eq.3}
\end{equation}

To achieve (\ref{Eq.3}), we need to find out the
output variable with the relative order of 1 to system (\ref{Eq.1}), supposing to be $y_m$, then we have:
\begin{equation}
  \Dot{y}_m=L_r{h}_m(\bm{x}) +
  L_s{L_{r}^0}{h}_m(\bm{x})u \, ,
  \label{Eq.4}
\end{equation}
where $L_{(\cdot)}$ is the Lie derivative operator \cite{Reithmeier1990Nonlinear}. To this end, the Brunovsky system can be constructed based on (\ref{Eq.2}) and (\ref{Eq.4}) as
\begin{equation}
  \Dot{\bm{I}} = \bm{AI} + \bm{B}v \, ,
  \label{Eq.5}
\end{equation}
where
\begin{equation*}
  \bm{I}=\begin{bmatrix} I_1\\I_2\\\vdots\\I_m \end{bmatrix}, \quad 
  \bm{A}=\begin{bmatrix}0&1&0&\cdots&0\\0& 0& 1&\cdots& 0\\ \vdots&
  \vdots& & \vdots& \vdots\\0& 0& 0& \cdots& 1\\0& 0& 0& \cdots&
  0 \end{bmatrix}, \quad \bm{B}=\begin{bmatrix}0\\0\\
  \vdots\\0\\1 \end{bmatrix} \, .
\end{equation*}
From (\ref{Eq.4}) and (\ref{Eq.5}), $v$ can be expressed as
\begin{equation}
  v = L_{r}{h}_m(\bm{x}) +
  L_{s}{L_{r}^0} {h}_m(\bm{x})u-\Dot{y}^{*}_{m} \, . 
  \label{Eq.6}
\end{equation}

Since the Brunovsky system (\ref{Eq.5}) is a completely controllable linear system, the linear control theory, such as linear quadratic optimal control \cite{Perez2018Linear} can be applied to obtain $v$, yielding:
\begin{equation}
  v=\sum_{k=1}^m-k_iI_i \, .
  \label{Eq.7}
\end{equation}

Using (2), (6) and (7), the control strategy of the nonlinear system (\ref{Eq.1}) can be derived as
\begin{equation}
  u=\frac{-\sum_{i=1}^m k_i(y_i-y^{*}_{i})-L_{{r}}h_m(\bm{x}) +
    \Dot{y}^{*}_{m}}{L_{{s}}L_{{r}}^0h_m(\bm{x})} \, .
  \label{Eq.8}
\end{equation}
The interested readers can find more details on the effectiveness of OHFT in \cite{Hui2008Nonlinear}.

\subsection{System Description}
In this section, the WTG model, the power grid primary frequency regulation model, the conventional VIC scheme and its impacts on driven-train torsional oscillation of WTG are presented and analyzed.

\subsubsection{Drive-train of WTG}
The two-mass drive-train model of WTG can be expressed as \cite{MaHybrid}
\begin{equation}
  \label{Eq.9}
  \begin{aligned}
    2H_t\frac{d\omega_t}{dt} &= T_m-K_{sh}\theta_{sh}-D_{sh}(\omega_t-\omega_g) \, , \\
    2H_g\frac{d \omega_g}{dt} &= K_{sh}\theta_{sh} + D_{sh}(\omega_t-\omega_g)-T_e \, , \\
    \frac{d\theta_{sh}}{dt} &= \omega_B(\omega_t-\omega_g) \, ,
  \end{aligned}
\end{equation}
where $H_t$ and $H_g$ are the inertia constants of the wind turbine and generator in seconds, respectively;  $\omega_t$ and $\omega_g$ denote the speeds of wind turbine and generator rotor in radians/s, respectively.
$K_{sh}$ and $D_{sh}$ represent the stiffness coefficient and the damping coefficient of the shaft, respectively. $P_m$, $P_e$ and $\theta_{sh}$ are the mechanical power, the total output active power of DFIG and the torsional angle between the wind turbine and generator rotor, respectively.

According to the law of aerodynamics, the mechanical power extracted by the WTG can be calculated by \cite{Bhende2016Frequency}:
\begin{equation}
  \label{Eq.10}
  \begin{aligned}
    P_m&=\frac{\pi \rho R^2V_w^3C_p(\lambda,\beta)}{2P_B} \, , \\
    \lambda&=\frac{R\omega_t\omega_B}{V_w} \, , \\
    C_p(\lambda,\beta)&=0.5176(\frac{116}{\lambda_i} -
    0.4\beta-5)e^{\frac{-21}{\lambda_i}} + 0.0068\lambda \, , \\
    \frac{1}{\lambda_i}&=\frac{1}{\lambda+0.08\beta}-\frac{0.035}{\beta^3+1} \, , 
  \end{aligned}
\end{equation}
where $\rho$, $V_w$, $P_m$, $\lambda$, $R$, $C_p(\lambda,\beta)$ and $P_B$ are the air density, wind speed, mechanical power, tip-speed ratio, rotor radius, power coefficient and WTG rated power, respectively.

\subsubsection{Modeling of WTG active power control loop}
According to \cite{Rahimi2018Improvement}, the control loop of WTG active power can be modeled as a low pass transfer function, i.e.,
\begin{equation}
  \frac{d P_e}{dt} = a_P (P_{\rm ref} - P_e) \, ,
  \label{Eq.11}
\end{equation}
where $a_P$ is the bandwith of WTG stator active power control loop; $P_{\rm ref}$ represents the reference electromagnetic power of WTG, which is determined by the MPP tracing strategy \cite{Chen2017} given by
\begin{equation}
  P_{\rm ref} = P_{\rm MPP} = k_{\rm opt} \omega_g^3 \, ,
  \label{Eq.12}
\end{equation}
where $k_{\rm opt}$ is the MPP tracing coefficient.  

\subsubsection{Conventional VIC of WTG}
The conventional VIC for WTG is developed to support the power grid frequency and it determines the reference electromagnetic power increment delivered from WTG to the grid, $P_{\rm vir}$. Formally, we have
\begin{equation}
  P_{\rm vir} = - k_{P_{\rm vir}}\Delta \omega -
  k_{D_{\rm vir}}\frac{d\Delta \omega}{dt} \, ,
  \label{Eq.13}
\end{equation}
where  $\Delta \omega = \omega-\omega_s$; $\omega$ and $\omega_s$ denote the angular
velocity of power grid and the synchronous angular velocity; $k_{\rm P_{\rm vir}}$ and $k_{D_{\rm vir}}$ are the positive coefficients of the proportional and differential elements, respectively. As a result, the actual reference electromagnetic power of WTG equals to:
\begin{equation}
  P_{\rm ref} = P_{\rm MPP} + P_{\rm vir} \, .
  \label{Eq.14}
\end{equation}

\subsubsection{Equivalent model of the grid}

\begin{figure}
  \centering
  \includegraphics[scale=0.25]{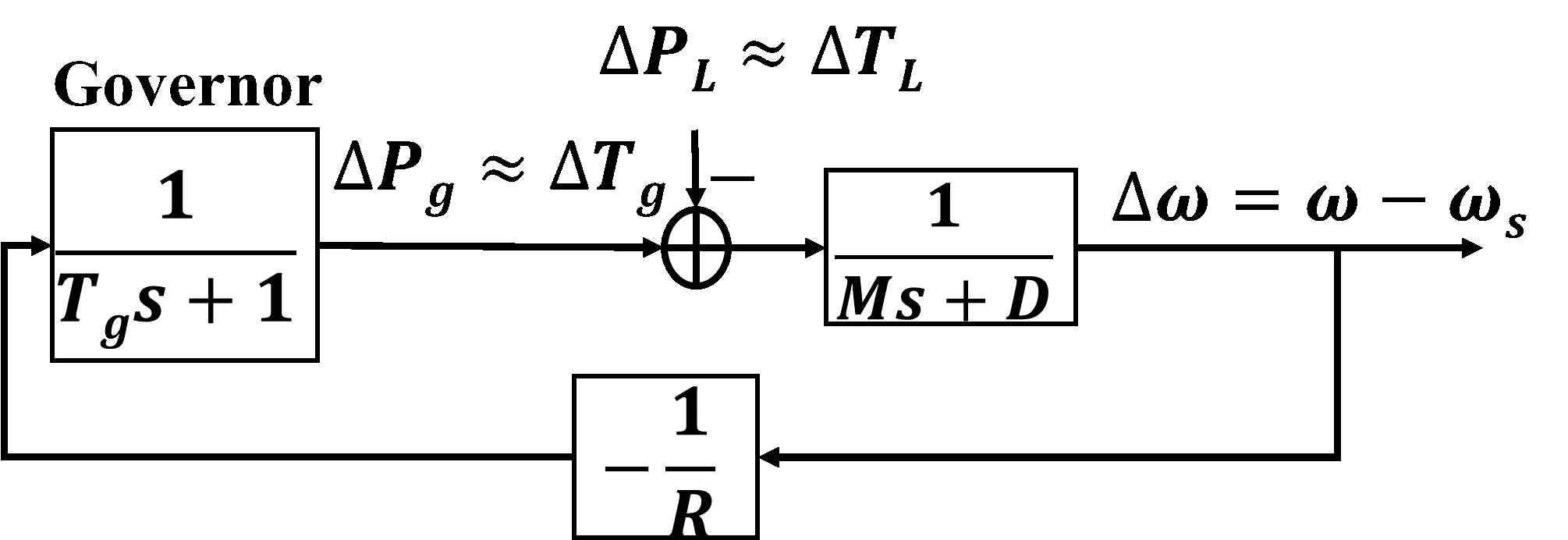}
  \caption{Block diagram of power grid primary frequency regulation.}
  \label{Fig.1}
\end{figure}

For the purposes of our analysis, which is focused on the WTG dynamic, the grid can be modeled for simplicity but without loss of generality as an equivalent synchronous generator plus a load \cite{Grigsby2015Power}.  The primary frequency regulation of the power grid is described by the block diagram shown in Fig.~\ref{Fig.1}, where $M$ is the equivalent inertia constant of the grid; $D$ is the frequency coefficient of load that is expressed as percent change in load divided by percent change in frequency; $\Delta P_L$ denotes the load variation in the power grid; $\Delta P_{g}$ is the increment of synchronous generator output power; $R$ is the equivalent droop coefficient and the governor of the equivalent synchronous generator is treated as an inertia element with time constant $T_g$.  Therefore, we have:
\begin{equation}
  \frac{d\Delta P_g}{dt}=-\frac{1}{RT_g} \Delta \omega -
  \frac{1}{T_g}\Delta P_g \, .
  \label{Eq.15}
\end{equation}

As the fluctuation of power grid frequency is small, the corresponding electromagnetic torque approximately equals to the electromagnetic power. Then, the dynamic equation of the synchronous generator rotor is approximated as:
\begin{equation}
  M\frac{d \Delta \omega}{dt}= \Delta P_{\rm tot} -D  \Delta \omega \, ,
  \label{Eq.16}
\end{equation}
where $\Delta P_{\rm tot} = \Delta P_g+P_e-P_{e0} - \Delta P_L$ and $P_{e0}$ is the initial output active power of WTG.

\subsubsection{Low frequency drive-train torsional oscillations caused by the VIC}

The transfer function from the electromagnetic torque increment of WTG, $\Delta T_e$, to the torsional angle variation, $\Delta \theta_{sh}$, can be derived from (\ref{Eq.9}). At the initial moment of frequency event, $\Delta T_e=\frac{P_{\rm vir}}{\omega_{g0}}$ holds, yielding:
\begin{equation}
  \Delta \theta_{sh}=\frac{\omega_B H_t}{2(H_t H_g)s^2 +
    a H_{tg}s+bH_{tg}}
  \frac{P_{\rm vir}}{\omega_{g0}} \, ,
  \label{Eq.17}
\end{equation}
where $H_{tg} =H_t+H_g$; $a = D_{sh} {\omega_B}$; $b=K_s{_h{\omega_B}}$ and $\omega_{g0}$ denotes the initial WTG rotor speed of MPP before VIC is activated. From (\ref{Eq.17}), we observe that the reference electromagnetic power increment requested by VIC may result in the low frequency oscillation if no damping control is provided. This is because there exists a torsional mode with the natural low frequency shown in (\ref{Eq.18}) and the damping coefficient $D_{sh}$ is usually small \cite{Rahimi2018Improvement}.
\begin{equation}
  \omega_n=\sqrt{\frac{\omega_BK_{sh}H_{tg}}{2H_tH_g}} \, .
  \label{Eq.18}
\end{equation}

\begin{figure}
  \centering
  \includegraphics[width=\linewidth]{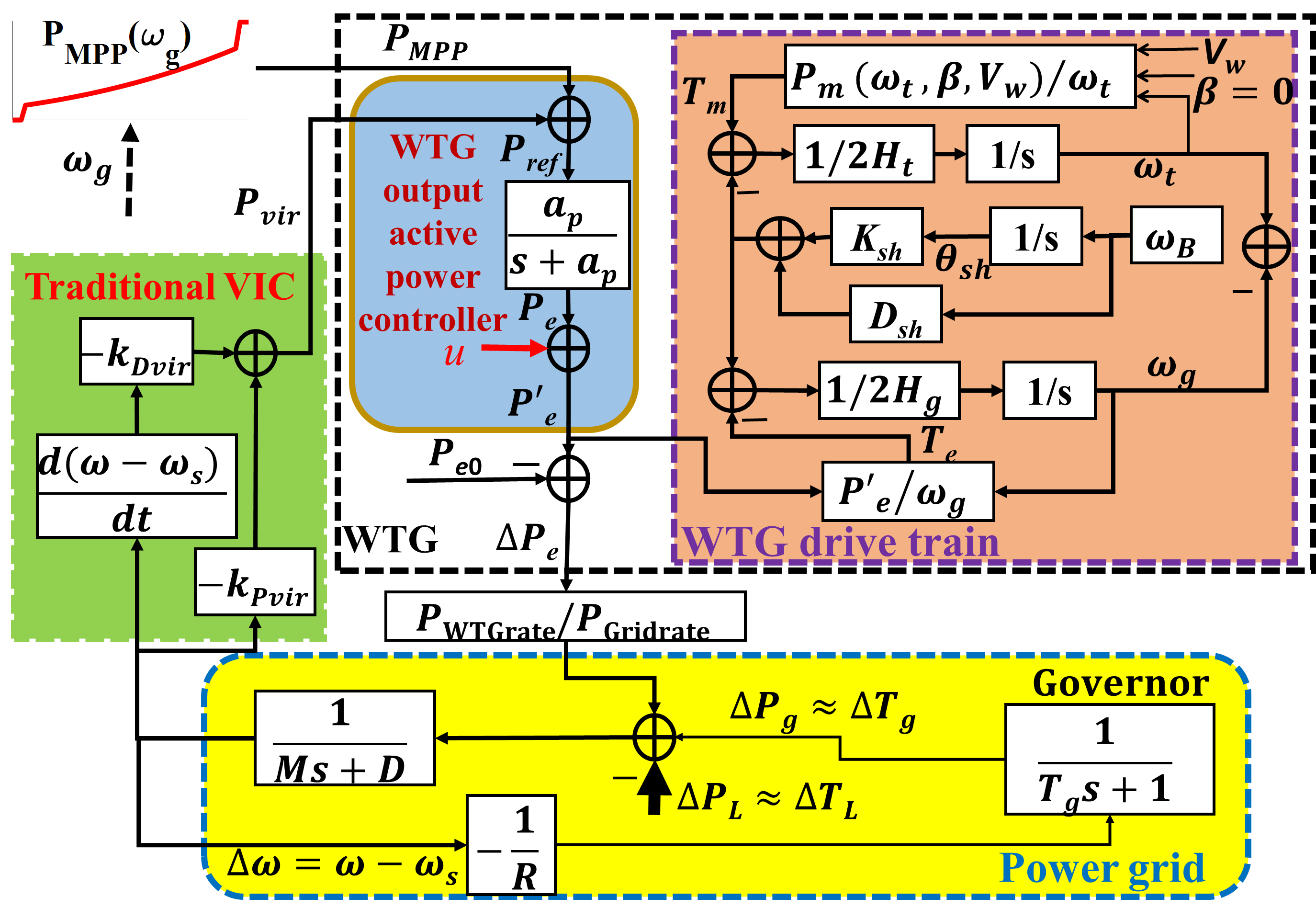}
  \caption{Block diagram of the whole system.}
  \label{Fig.2}
\end{figure}

The complete system block diagram is reported by Fig.~\ref{Fig.2}, where $u$ is the control input. The power grid and WTG are in pu by their own rated power, i.e., $P_{\rm Gridrate}$ and $P_{\rm WTGrate}$, respectively. In this paper, the nominal values of angular velocity with respect to power grid and WTG rotor are $120\pi$ rad/s and $120 \pi/p$ rad/s, respectively, where $p$ represents the number of WTG pole-pairs.

\subsection{Proposed VIC With Single WTG}

The state-space description of the whole system can be expressed in the form of (\ref{Eq.1}), where the state variables are $[\omega_t,\omega_g,\theta_{sh}, P_e, \Delta P_g, \omega]$.  This paper aims at designing a nonlinear VIC of WTG, which (i) suppresses low frequency drive-train torsional oscillations of WTG; (ii) prevents large frequency excursions; and (iii) facilitates the smooth and fast recovery of the kinetic energy of the WTG.

For the former two objectives, based on the aforementioned OHFT, we first choose $\omega_{tg}$ and $\Delta \omega$ as the output variables and derive the following Brunovsky system:
\begin{equation}
  \begin{bmatrix}
    \Dot{\omega}_{tg} \\
    \Delta \Dot{\omega}
  \end{bmatrix} =
  \begin{bmatrix}
    0&1\\0&0
  \end{bmatrix}
  \begin{bmatrix}
    \omega_{tg} \\
    \Delta \omega
  \end{bmatrix}+
  \begin{bmatrix}
    0\\1
  \end{bmatrix} v \, ,
  \label{Eq.19}
\end{equation}
where $\omega_{tg} = \omega_t-\omega_g$. Note that $\Delta \omega$ is the output with relative order of 1 to the studied system depicted by Fig.~\ref{Fig.2}. By substituting $u$ into (\ref{Eq.16}) yields
\begin{equation}
  M\frac{d \Delta \omega}{dt} = \Delta P_{\rm tot} - D\Delta \omega + u \, .
  \label{Eq.20}
\end{equation}

The change rate of the WTG torsional angle $d\theta_{sh}/{dt}$ and the power grid angular frequency $\omega$ are required to be close to 0 and 1 pu as possible, respectively. These objectives can be achieved by the linear quadratic optimal control, whose performance indicator about system \eqref{Eq.19} is constituted as follows:
\begin{equation}
  J=\int_0^\infty \bm{I^TQI}+\alpha v^2dt \, ,
  \label{Eq.21}
\end{equation}
where $\bm{Q}$ is a symmetric positive definite matrix; coefficient $\alpha$ is a positive real number and $\bm{I}=[\omega_{tg}, \Delta \omega]^T$.
According to the control strategy (\ref{Eq.8}) designed by means of OHFT, we have the nonlinear state feedback control strategy of the targeted system (\ref{Eq.19}) as follows:
\begin{equation}
  \begin{split}
    u=-Mk_1\omega_{tg} -
    (Mk_2-D)\Delta \omega - \Delta P_{\rm tot} \, .
  \end{split}
  \label{Eq.22}
\end{equation}

By converting $u$ to a quantity of the conventional VIC output interface into $u'$, we have
\begin{equation}
  u' =u+\frac{1}{a_P}\frac{du}{dt} \, .
  \label{Eq.23}
\end{equation}

Combining (\ref{Eq.23}) with the conventional VIC expressed by (\ref{Eq.13}), a nonlinear VIC can be obtained as:
\begin{equation}
  \label{Eq.24}
  P'_{\rm vir}= P_{\rm vir} + u'  \, .
\end{equation}

\begin{figure}
  \centering
  \subfloat[]{
    \begin{minipage}[t]{0.49\linewidth}
      \centering
      \includegraphics[width=\columnwidth]{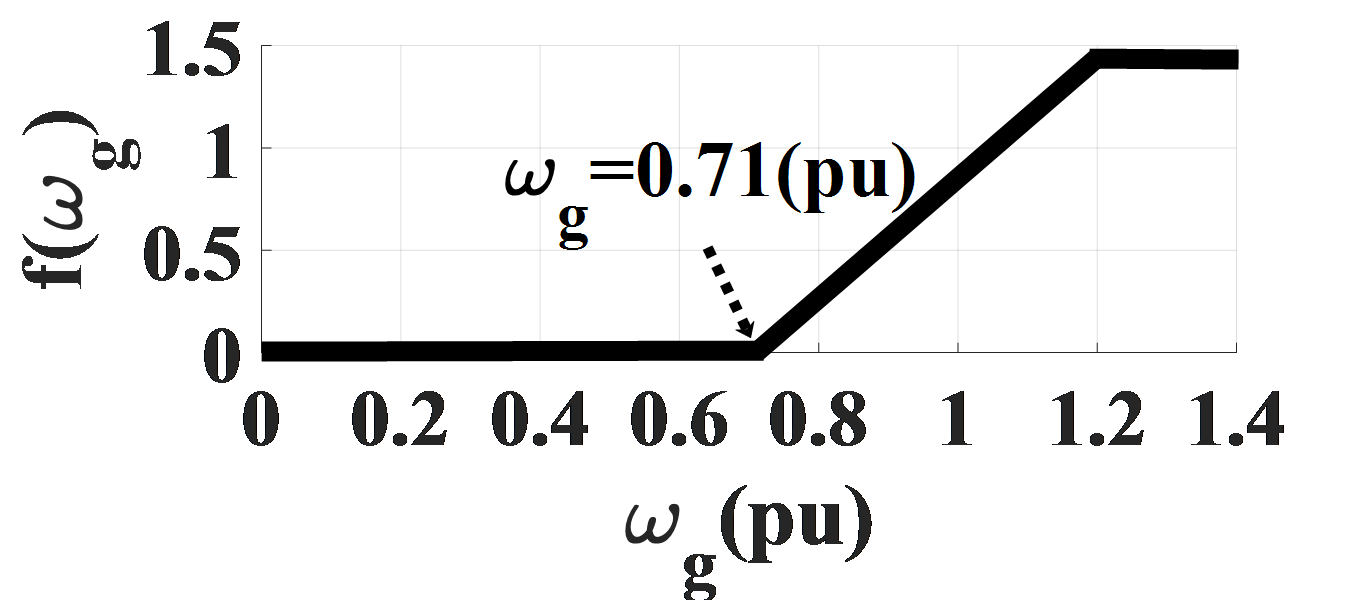}
    \end{minipage}
  }
  \subfloat[]{
    \begin{minipage}[t]{0.49\linewidth}
      \centering
      \includegraphics[width=\columnwidth]{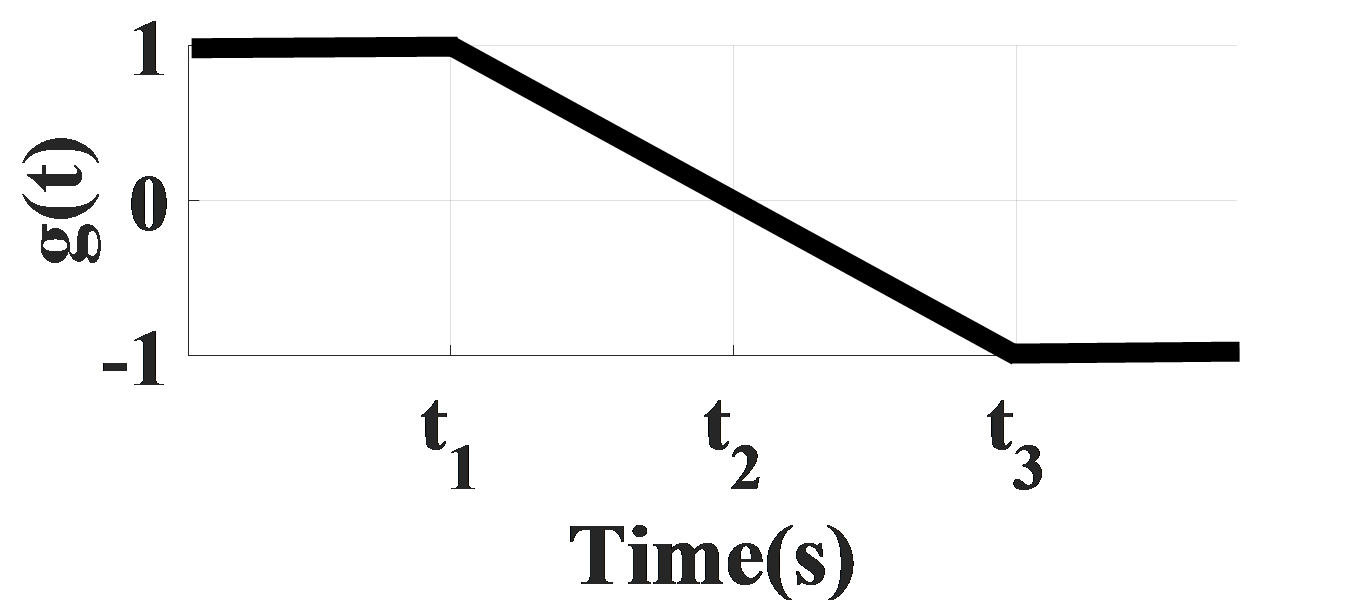}
    \end{minipage}
  }
  \centering
  \caption{(a) Function of rotor speed $f(\omega_g)$; (b) Function
    of time $g(t)$.}
  \label{Fig.3}
\end{figure}

To further satisfy the reference electromagnetic power increment required by VIC, $P'_{\rm vir}$ can be adjusted following the WTG rotor speed $\omega_g$. In other words, VIC requires more power increment at the high rotor speed when there is enough kinetic energy that can be released, and vice versa. Thus, we multiply the result of VIC in (\ref{Eq.24}) by a function of rotor speed, $f(\omega_g)$, yielding:
\begin{equation}
  P''_{\rm vir}= P'_{\rm vir} \, f(\omega_g) \, ,
  \label{Eq.25}
\end{equation}
where $f(\omega_g)$ is shown in Fig.~\ref{Fig.3}(a).

To achieve the last goal, namely facilitating smooth
and fast recovery of the kinetic energy of WTGs, a function of time $g (t)$ shown in Fig.~3(b) is multiplied by (\ref{Eq.25}). A positive $g(t)$ promotes WTG to provide active power support for power grid while a negative $g(t)$ is effective in the recovery of WTG rotor speed. Note that function $g(t)$ also allows the power for frequency support to vary in a smooth manner. This is beneficial to suppress the torsional oscillation of WTG drive train caused by the instantaneous variation of reference power, avoiding the secondary frequency dip.

Finally, the proposed VIC can be given by 
\begin{equation}
  \label{Eq.26}
  P'''_{\rm vir}=  P''_{\rm vir} \, g(t) = P'_{\rm vir} \, f(\omega_g) \, g(t) \, .
\end{equation}

\subsection{Extension to Multiple WTGs}
\begin{figure}
  \centering
  \includegraphics[width=\linewidth]{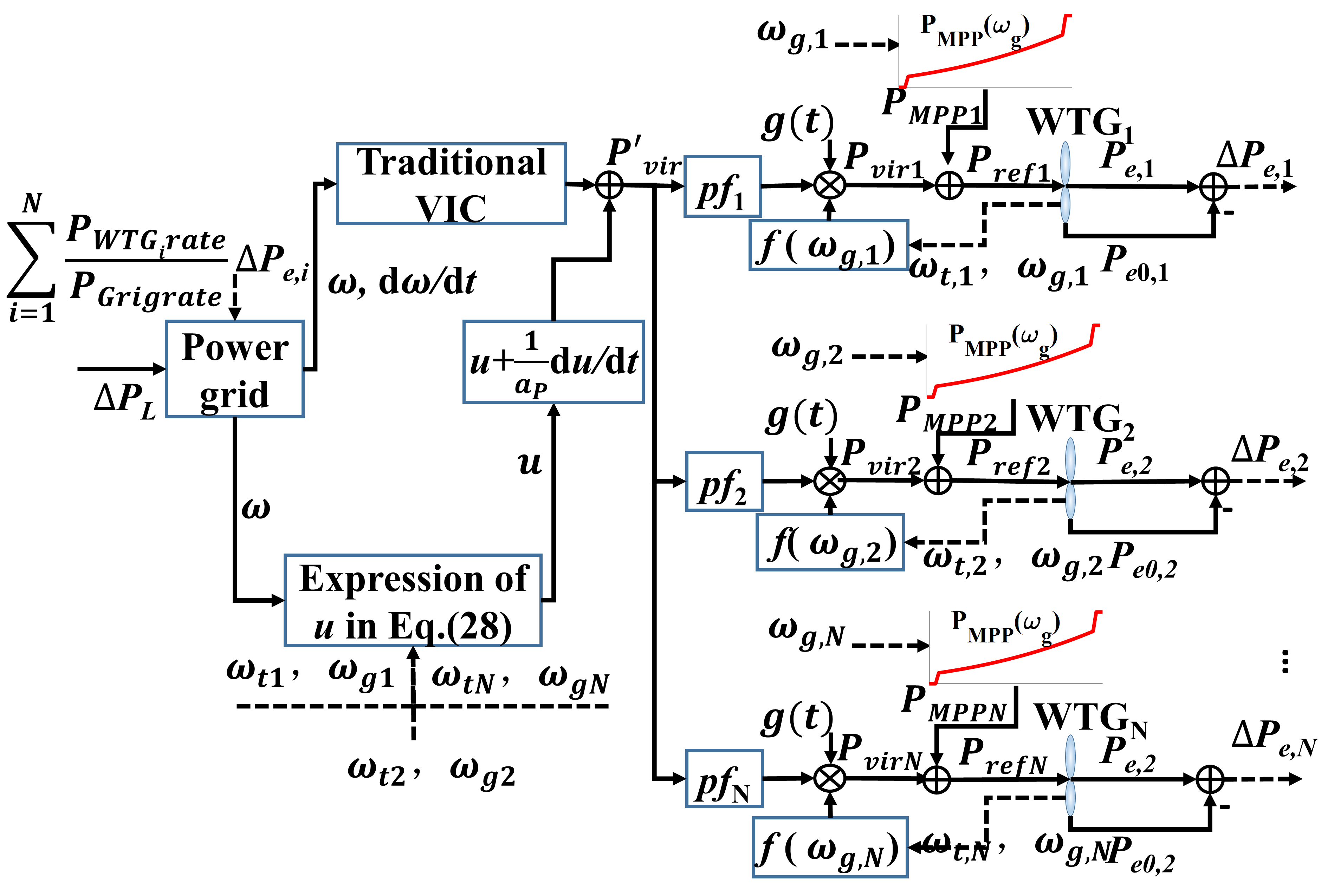}
  \caption{The block diagram of the proposed nonlinear VIC.}
  \label{Fig.4}
\end{figure}

The proposed nonlinear VIC controller is now extended to multiple WTGs as shown in Fig.~\ref{Fig.4}. For multiple WTGs, the Brunovsky system can be constructed as
\begin{equation}
  \begin{bmatrix}
    \Dot{\omega}_{tg,1}\\
    \Dot{\omega}_{tg,2}\\
    \vdots\\
    \Dot{\omega}_{tg, N}\\
    \Delta \Dot{\omega}
  \end{bmatrix} = \\
  \begin{bmatrix}
    0&1&0&\cdots&0\\
    0&   0&   1& \cdots & 0\\
    \vdots& \vdots&  &  \vdots&  \vdots\\
    0& 0& 0& \cdots&  1\\
    0& 0& 0& \cdots& 0
  \end{bmatrix}
  \begin{bmatrix}
    \omega_{tg,1} \\
    \omega_{tg,2} \\
    \vdots \\
    \omega_{tg,N} \\
    \Delta \omega
  \end{bmatrix} +
  \begin{bmatrix}
    0\\0\\ \vdots\\0\\1
  \end{bmatrix} v \, ,  
  \label{Eq.27}
\end{equation}
where the subscript $i$ denotes the quantities with respect to the $i$th WTG and $N$ is the number of WTGs.

Similar to the design procedure of single-WTG VIC, the signal $P'''_{\rm vir, tot}$ is obtained as:
\begin{equation}
  \label{Eq.28}
  \begin{aligned}
    P'''_{\rm vir, tot} =& \sum_{i=1}^N P'_{{\rm vir}} \cdot {\rm pf}_i \cdot f(\omega_{g,i})\cdot g(t) \, ,
  \end{aligned} 
\end{equation}
where ${\rm pf}_i$, $i=1,2,\dots,N$ represents the participating factor of the $i$th WTG; the control input $u$ in $P'_{vir}$ equals to $-\sum_{i=1}^N {M{k_i}\omega_{tg,i}-(M\;k_{N+1}-D)\Delta \omega -\Delta P_{tot}}$; for multiple WTGs, $P_e=\sum_{i=1}^N P_{e,i}$ and $P_{e0}=\sum_{i=1}^N P_{{e0},i}$, where $P_{{e0},i}$ and $P_{e,i}$ denote the initial and real-time output active power of the $i$th WTG, respectively; note that all power should be converted into quantities in pu by nominal power of the power grid; $k_1,k_2, \dots,k_{N+1}$ represent the state feedback coefficients when applying linear quadratic optimal control for the system (\ref{Eq.27}) to obtain control quantity $v$. The participating factor of the $i$th WTG, ${\rm pf}_i$, determines the power support from each WTG and it is correlated with the WTG initial operation point. Formally, we have
\begin{equation}
  \label{Eq.29}
  {\rm pf}_i=\frac{P_{e0,i}}{\sum_{i=1}^k P_{e0,i}}
  \qquad i=1,2,\dots, N \, ,
\end{equation}
where the initial active power output of the $i$th WTG, $P_{e0,i}$, can be calculated based on the power flow solution at the moment of contingency. 

The block diagram of the proposed nonlinear VIC is displayed in Fig. \ref{Fig.4}. Similar to the proposed VIC for single WTG, the VIC provided by \eqref{Eq.28} also has several advantages, including 1) effectively suppressing the low frequency drive-train torsional oscillation of each WTG, 2) improving the power grid frequency nadir through the coordination of all WTGs, 3) achieving smooth and fast recovery of each WTG rotor speed to the corresponding original MPP point before the disturbance and preventing the secondary frequency dip. These benefits will be demonstrated in the simulation results section.

\begin{table}[h]
	\centering
	\caption{Parameter Settings of the Test System}
	\label{Table.I}
	\begin{tabular}{cc|cc}
		\hline
		\multicolumn{4}{c}{\textbf{Equivalent power grid}} \\ \hline
		$M$ & $4.584$ s & $D$ & $1$ \\
		$T_g$ & $1.2$ s & $R$ & $0.03$ \\
		$\Delta P_L$ & $0.2$ pu & $\omega_B$ & $377$ rad/s \\
		$P_{\rm Gridrate}$ & $3$ MW \\
		\hline \hline
		\multicolumn{4}{c}{\textbf{WTG}}\\ \hline
		$P_{\rm WTGrate}$ & $1.5$ MW & $\omega_B$ & $377/3$ rad/s \\
		$H_g$ & $0.685$ s & $H_t$ & $4.32$ s \\
		$D_{\rm sh}$ & $1.5$ & $K_{sh}$ & $1.1$ pu/rad \\
		$k_{\rm opt}$ & $0.4425$ & $\omega^{\min}_g$ & $0.71$ pu \\
		$a_P$ & $31.4$ rad/s & $p$ & $3$ \\
		\hline \hline
		\multicolumn{4}{c}{\textbf{Conventional VIC}} \\ \hline
		$k_{P_{\rm vir}}$ & $7$ & $k_{D_{\rm vir}}$ & $2$ \\ \hline \hline
		\multicolumn{4}{c}{\textbf{Function} $g(t)$} \\ \hline
		$t_1$ & $25$ s & $t_2$ & $52$ s \\
		$t_3$ & $79$ s \\ \hline
	\end{tabular} 
\end{table}

\section{Simulation results}
\begin{figure}
  \centering
  \includegraphics[scale=0.25]{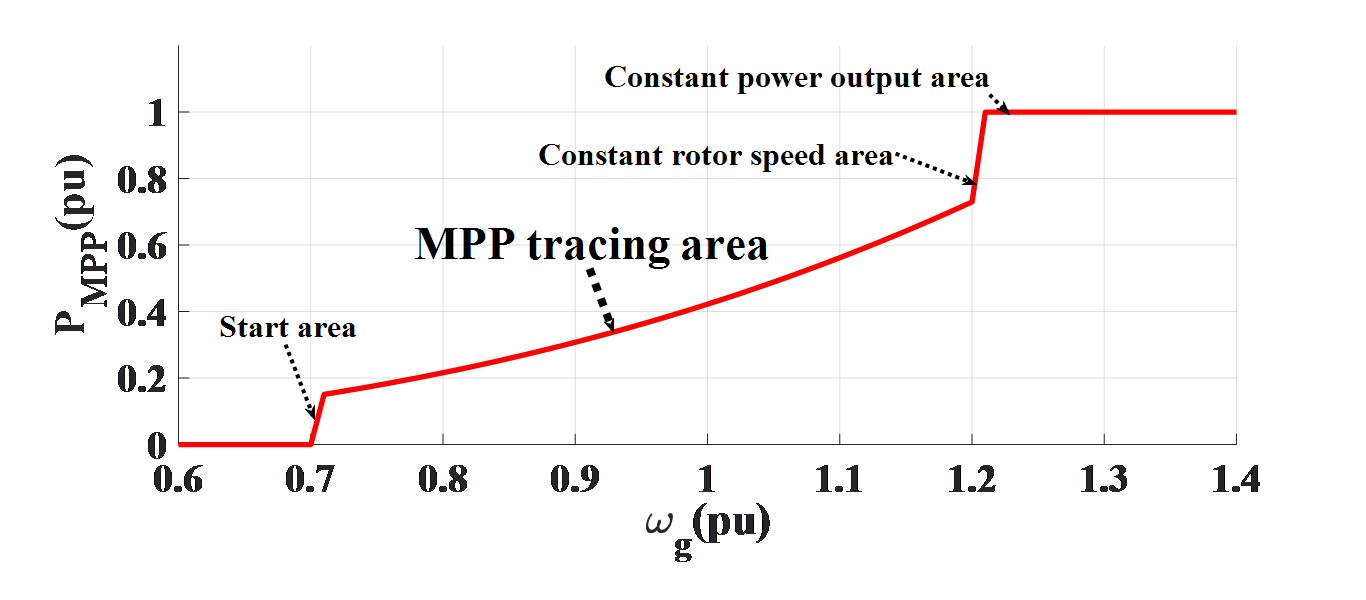}
  \caption{The MPP tracing curve of WTG.}
  \label{Fig.5}
\end{figure}

In this section, simulations are carried out to demonstrate the advantages of the proposed VIC. The conventional constant-coefficient VIC presented in \eqref{Eq.13} and the VIC-I defined by:
\begin{equation}
  P''''_{\rm vir}= P_{\rm vir} \, g(t)
  \label{Eq.30}
\end{equation}
are used for comparisons. Note that the VIC-I is different from that in \cite{Garmroodi2017Frequency} with time-varying droop coefficient, i.e., the proportional coefficient. Here, both the proportional and differential coefficients are time-varying in VIC-I. The parameters of equivalent power grid, WTG, conventional VIC and $g(t)$ are reported in Table \ref{Table.I}; the nominal power grid frequency is 60 Hz. It is worth pointing that the active power output increment of WTG, $\Delta P_e$, is the sum of reference electromagnetic power increment according to the MPP tracing strategy shown in (\ref{Eq.12}) and the extra active power required by the VIC, i.e., $\Delta P_e=\Delta P_{\rm MPP}+P_{\rm vir}$. The MPP tracing curve of WTG is presented by Fig.~\ref{Fig.5} and this paper concentrates on the MPP tracing area.

Following \cite{Danny2017Fast}, the equivalent grid model described in Section II is adopted in simulations.  The conventional VIC–based system, i.e., the open loop system shown in Fig.~\ref{Fig.2}, the modified VIC-I based system according to (\ref{Eq.30}) and the proposed VIC-based system shown in Fig.~\ref{Fig.4} are implemented in Matlab/Simulink.  Section III-A is used to demonstrate the advantages of proposed VIC with single WTG and its performance under different wind speed scenarios is investigated in Section III-B. Results on multiple WTGs are shown in Section III-C. For all simulations, a load contingency with 0.2 pu power increment occurs at 20 s in the power grid and the conventional VIC maintains for a preset period of 20 s.

\subsection{VIC for Single WTG}

\begin{figure}
  \centering
  \includegraphics[scale=0.22]{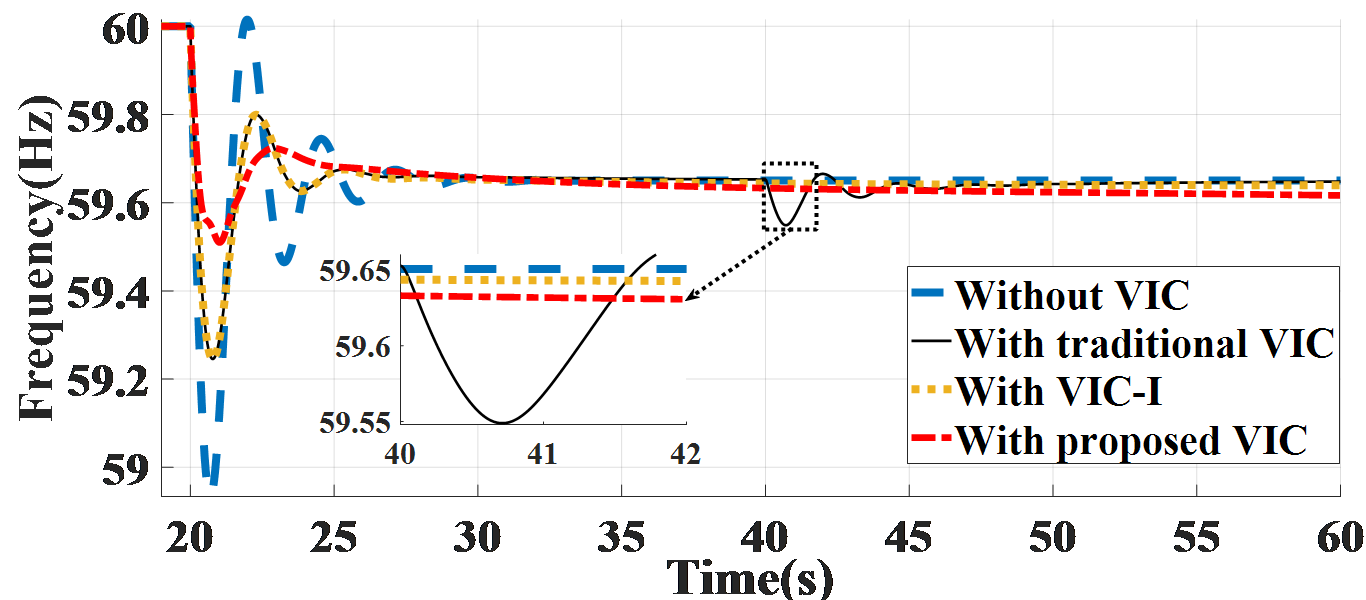}
  \caption{Frequency response with different VICs.}
  \label{Fig.6}
\end{figure}
\begin{figure}
  \centering
  \includegraphics[scale=0.22]{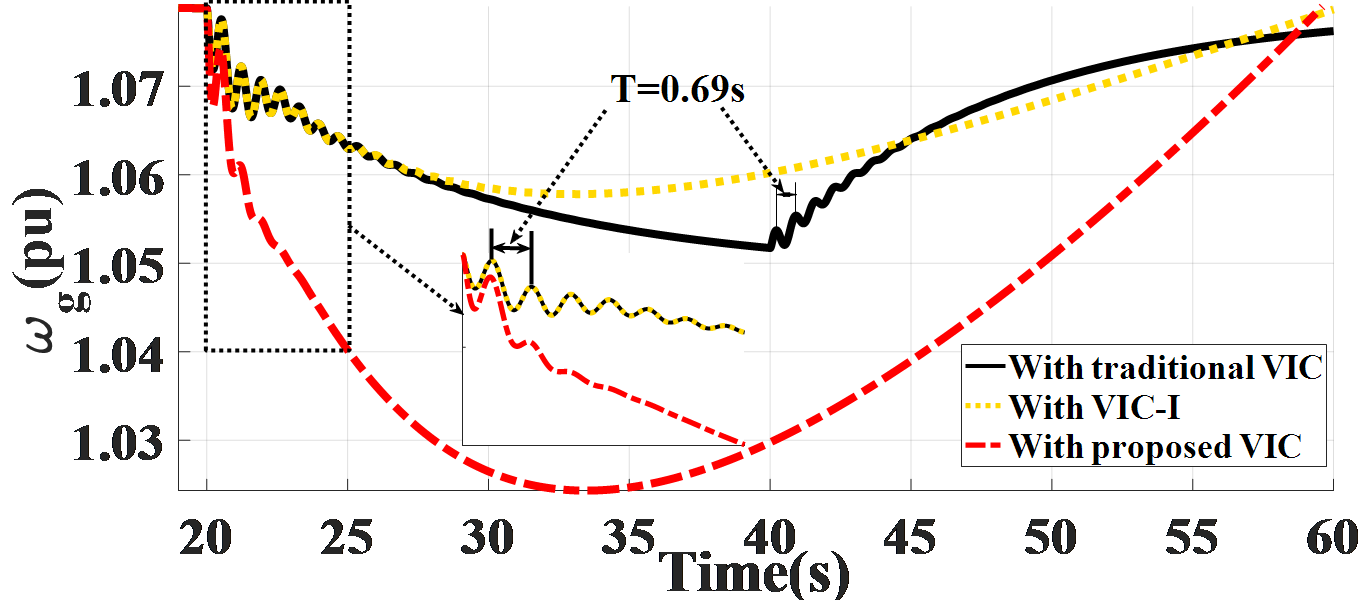}
  \caption{Rotor speed responses of WTG with different VICs.}
  \label{Fig.7}
\end{figure}
\begin{figure}
  \centering
  \includegraphics[scale=0.25]{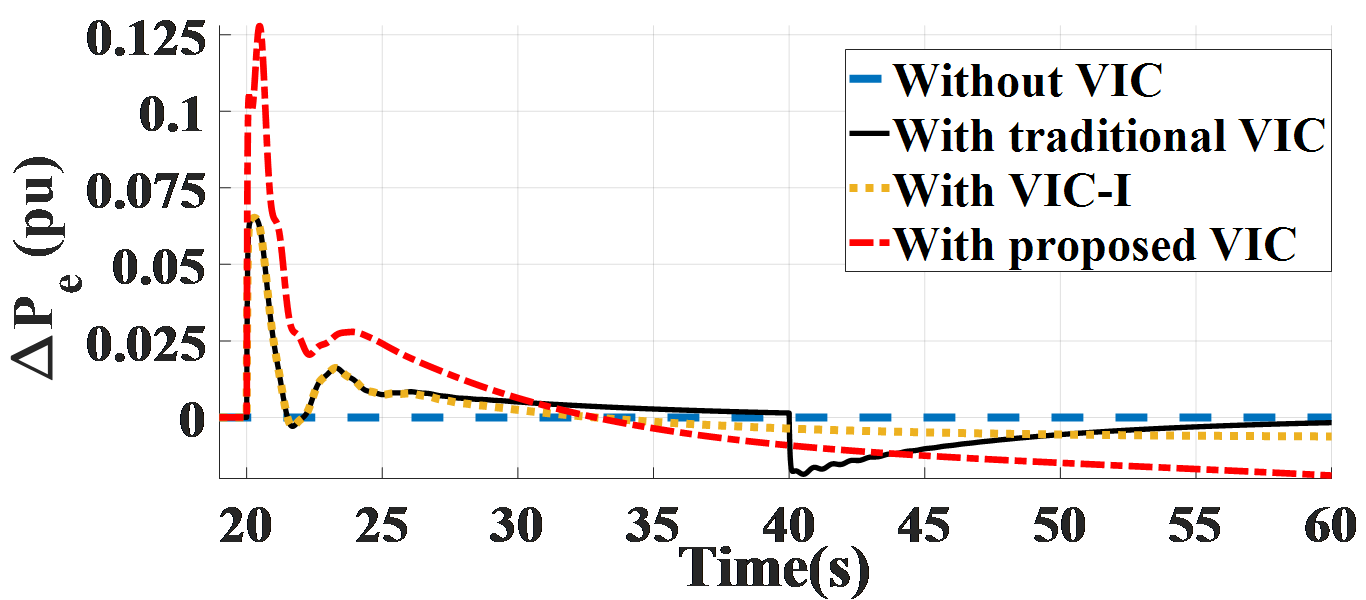}
  \caption{Active power output increments of WTG with different VICs.}
  \label{Fig.8}
\end{figure}

The wind speed is assumed to be 10.8 m/s and the state feedback coefficients of the linear quadratic optimal control for the system (\ref{Eq.19}) are $k_1$=2.6458 and $k_2$=2.5083. Fig.~\ref{Fig.6} presents the frequency responses with different VICs and the rotor speed response of WTG is shown in Fig. \ref{Fig.7}. Fig.~\ref{Fig.8} reports the active power output increment of WTG. Note that the wind power
penetration rate equals to
$33.3\% =(P_{\rm WTGrate} /(P_{\rm Gridrate} + P_{\rm WTGrate}))$.

According to the results, we can find that the frequency nadir obtained by the proposed VIC is much higher than the conventional VIC and improved VIC-I, see Fig.~\ref{Fig.6}. This is very important as the system frequency needs support for the first few seconds in the presence of contingency, especially for a system with high penetration of renewable energy and being vulnerable to large frequency decline. The reason is that the proposed VIC allows the WTG to reduce rotor speed as much as possible with a much higher speed than others, see Fig.~\ref{Fig.7}. This leads to much more active power increment from WTG, see Fig.~\ref{Fig.8}. It is interesting to observe from Fig.~\ref{Fig.6} that the power grid encountered a secondary frequency dip at 40s if the conventional VIC is used. This leads to a sudden reduction in the active power output of WTG, see the negative value of $\Delta P_e$ in Fig.~\ref{Fig.8}. By contrast, this is not an issue for the proposed VIC and the modified VIC-I as they can ensure the smooth exchange of power between WTG and power grid. As a result, the recovery of corresponding rotor speeds to MPP is smooth.

Fig. \ref{Fig.7} shows that both conventional VIC and modified VIC-I are subject to low frequency drive-train torsional oscillation with period 0.69s (the oscillation frequency is 1.45Hz). However, the proposed VIC can effective suppress that. Note that the torsional oscillation is a result from the sudden variation of WTG active power output as analyzed in (\ref{Eq.17}) and (\ref{Eq.18}). By observing the slopes of rotor speed responses in Fig.~\ref{Fig.7}, it can be concluded that the proposed VIC has higher recovery rate of WTG rotor kinetic energy than other two methods (note that the proposed method is the fastest one to recover rotor speed to the original MPP point before the disturbance and this would become more obvious in the results shown later). This is because after the main frequency support at about 32s, the proposed VIC prompts WTG to absorb power from grid at an earlier time in contrast to the conventional VIC. 

The comparison of the results above shows that the proposed VIC is effective in preventing large frequency excursions, maintaining smooth and fast recovery of the kinetic energy of WTG, and achieving the capability of suppressing WTG low frequency drive-train torsional oscillation.

\vspace{-0.3cm}
\subsection{VIC for Single WTG with Different Wind Speeds}
\begin{figure*}
  \centering
  \includegraphics[scale=0.32]{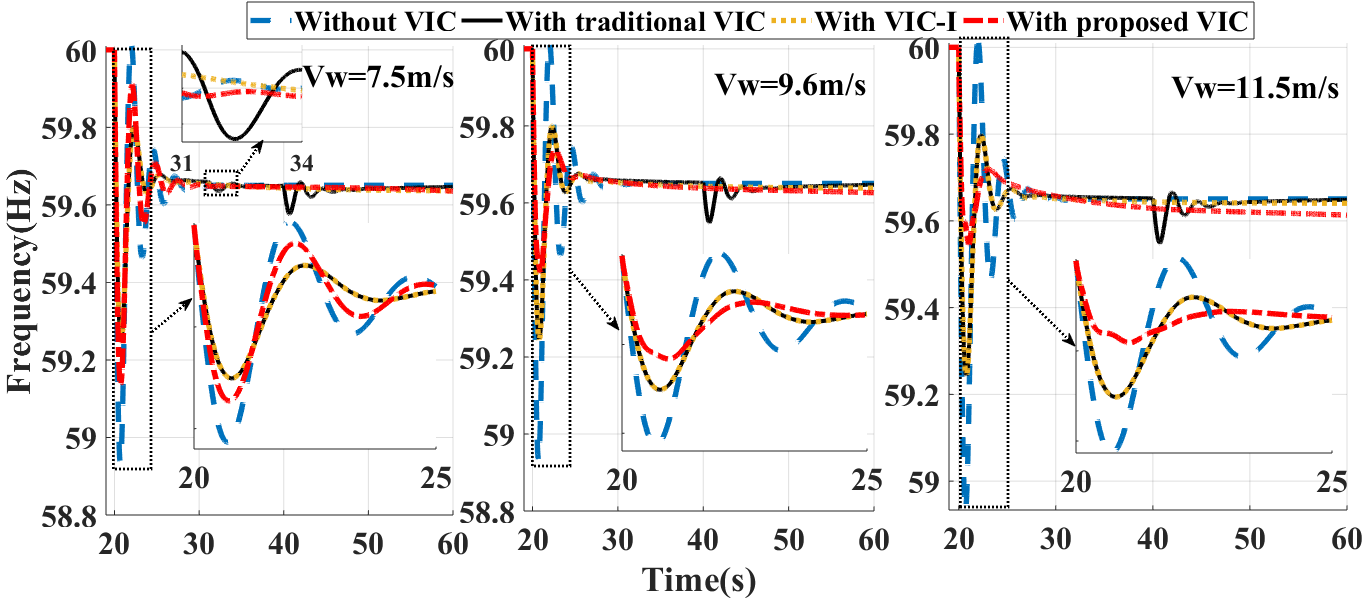}
  \caption{Frequency responses for different wind speeds}
  \label{Fig.9}
\end{figure*}

\begin{figure*}
  \centering
  \includegraphics[scale=0.32]{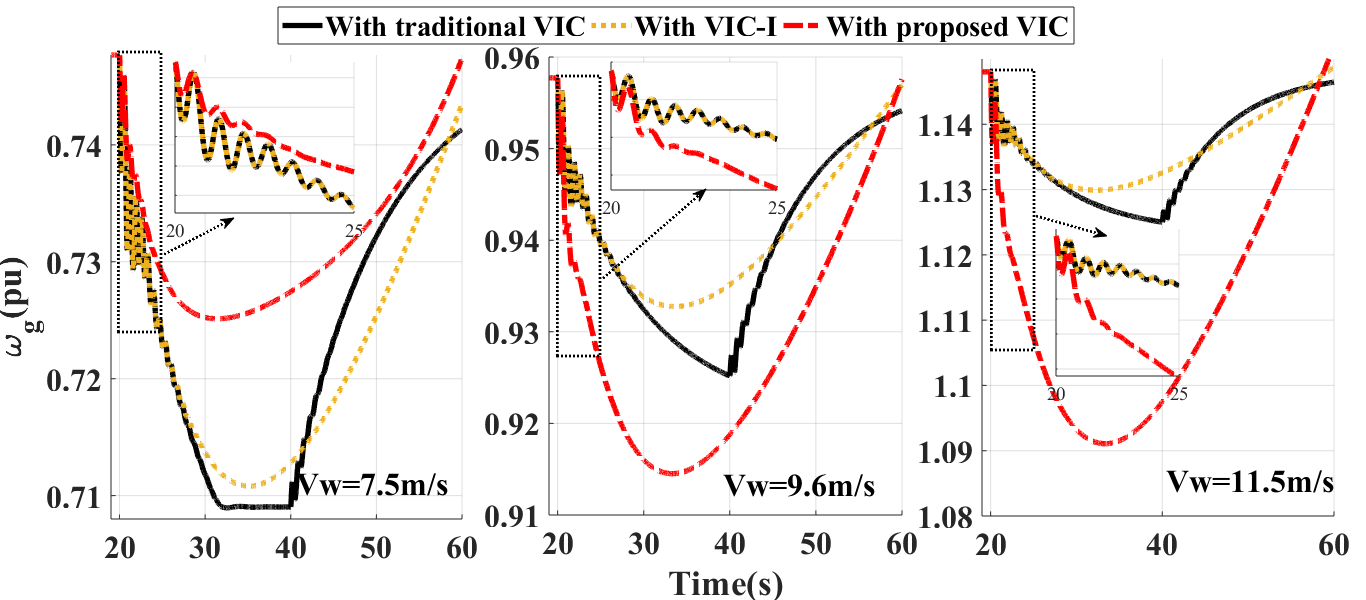}
  \caption{Rotor speed responses for different wind speeds}
  \label{Fig.10}
\end{figure*}

\begin{figure}
  \centering
  \includegraphics[scale=0.25]{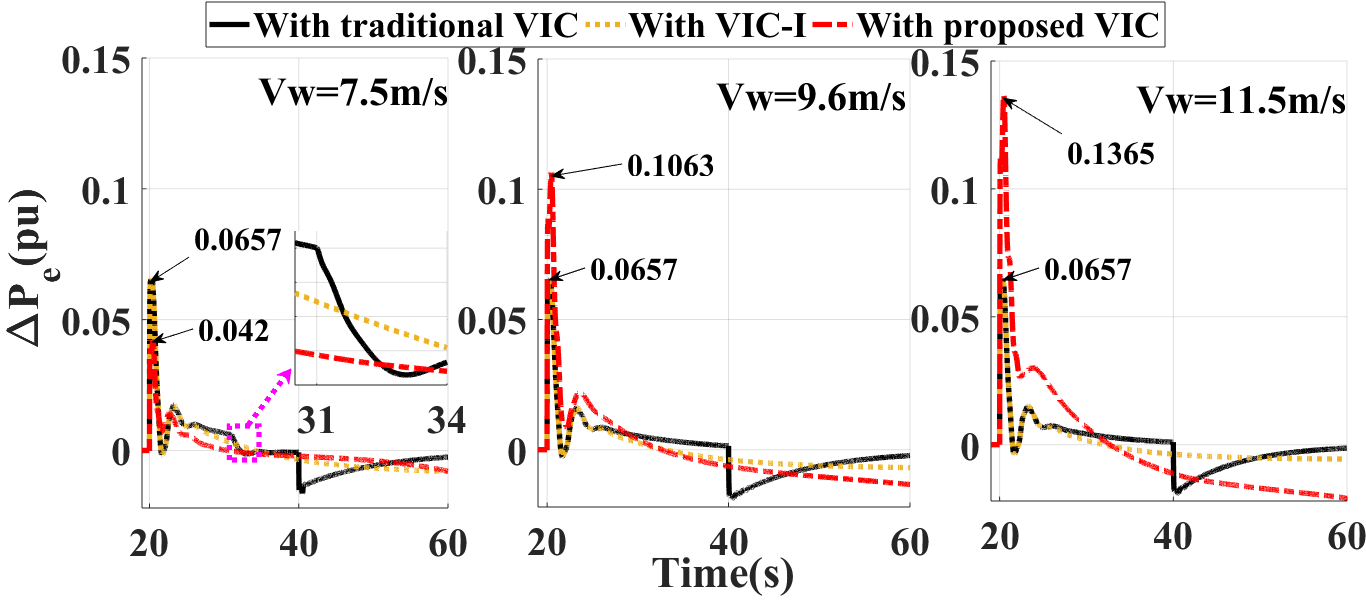}
  \caption{Active power increment of WTG for different wind speeds}
  \label{Fig.11}
\end{figure}

The proposed VIC is also tested under different wind speeds to demonstrate its robustness to various operating conditions. The following values for the wind speeds are tested: 7.5 m/s, 9.6 m/s and 11.5 m/s.  The frequency responses, WTG rotor speed responses and active power output increment of WTG are displayed in Fig.~\ref{Fig.9}, Fig.~\ref{Fig.10} and Fig.~\ref{Fig.11}, respectively.

For the low wind speed scenario, i.e., $7.5$ m/s, WTG with proposed VIC provides slightly less active power output increment than the other two methods in the initial stage of frequency regulation, see the first figure of Fig.~\ref{Fig.11}. This results in a slightly lower frequency nadir as shown in the first figure of Fig.~\ref{Fig.9}. It is for the purpose of preventing the
reference power of MPP-based tracing strategy from having a sudden dip to 0 as the rotor speed decreases below the minimal value. Note that the minimal value here is 0.71 pu.  While this case is encountered by the conventional VIC-based system, a sudden reduction in WTG active power output causes the secondary frequency dip from 31 to 34 s, see Fig.~\ref{Fig.9} and Fig.~\ref{Fig.11}. From Fig. \ref{Fig.10}, we can find that the other two methods are still subject to the WTG low frequency drive-train torsional oscillation while the proposed method can effectively suppress that. For higher wind speed scenarios, i.e., 9.6 m/s and 11.5 m/s, the proposed method achieves much better performances in terms of improving the frequency nadir and contributing to the smooth and fast recovery of the WTG rotor speed to MPP. Indeed, by comparing Fig.~\ref{Fig.7} with Fig.~\ref{Fig.10}, it can be found that the proposed method has more clear advantage of recovering rotor speed
to the original MPP point before the disturbance. It is also able to suppress the WTG low frequency drive-train torsional oscillation under different wind speed scenarios.

It is worth pointing out that the proposed VIC can automatically adjust the active power output increment according to the WTG operating point.  To elaborate on that, according to Fig.~\ref{Fig.11}, the active power output increment of WTG with the proposed VIC increases with the wind speed at the initial frequency regulation stage. The purpose is to prevent the excessive power demand because of the lack of kinetic energy stored by WTG rotor under low wind speed scenario. By contrast, it releases a lot of kinetic energy as soon as possible to support the power grid frequency under higher wind speed scenarios. The other two methods do not exhibit this characteristic as they lack of the adaptiveness.  This justifies why the frequency nadir with the proposed VIC is much higher than other two VICs under high wind speeds but it achieves slightly lower frequency nadir under low wind speed scenario.

\subsection{Proposed VIC with Multiple WTGs}

\begin{figure}
  \centering
  \includegraphics[scale=0.225]{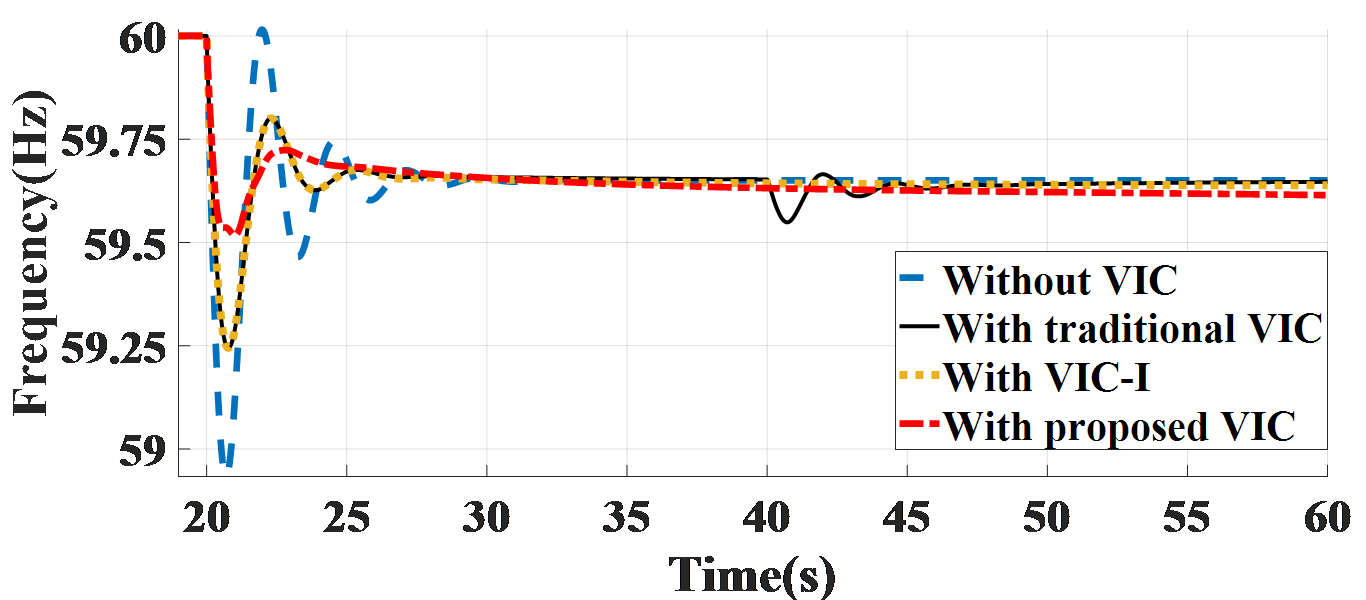}
  \caption{Frequency responses with multiple WTGs integrated grid.}
  \label{Fig.12}
\end{figure}

\begin{figure}
  \centering
  \includegraphics[scale=0.225]{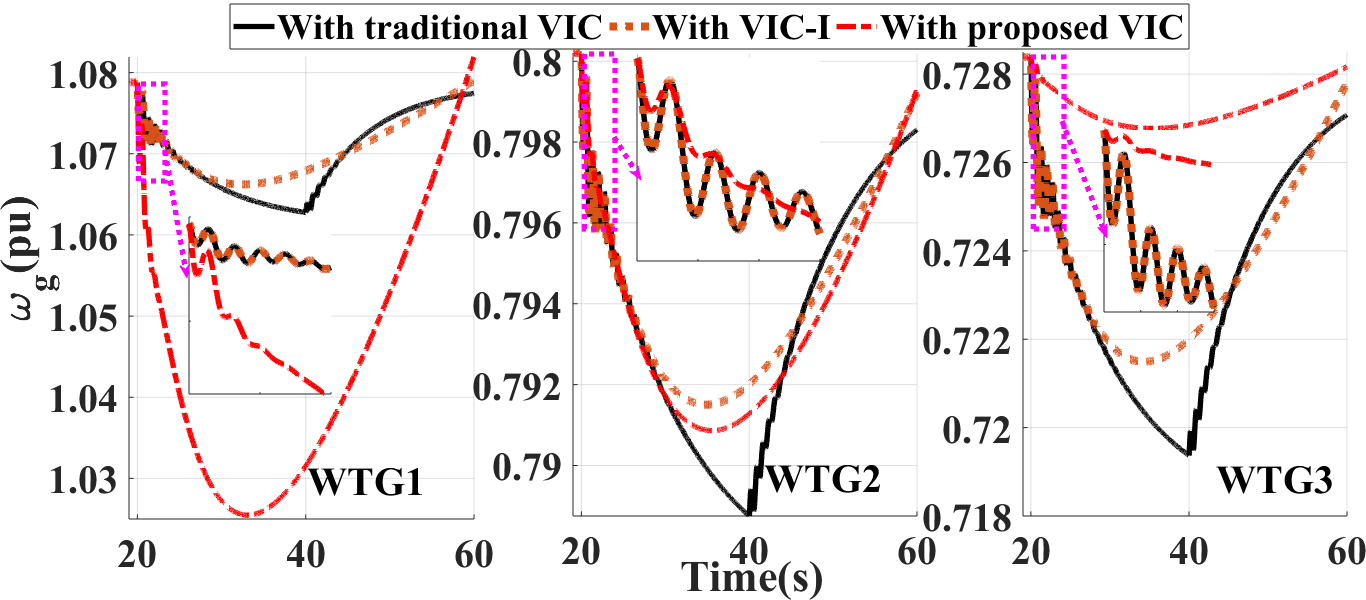}
  \caption{Rotor speed responses of WTGs.}
  \label{Fig.13}
\end{figure}

\begin{figure}
  \centering
  \includegraphics[scale=0.225]{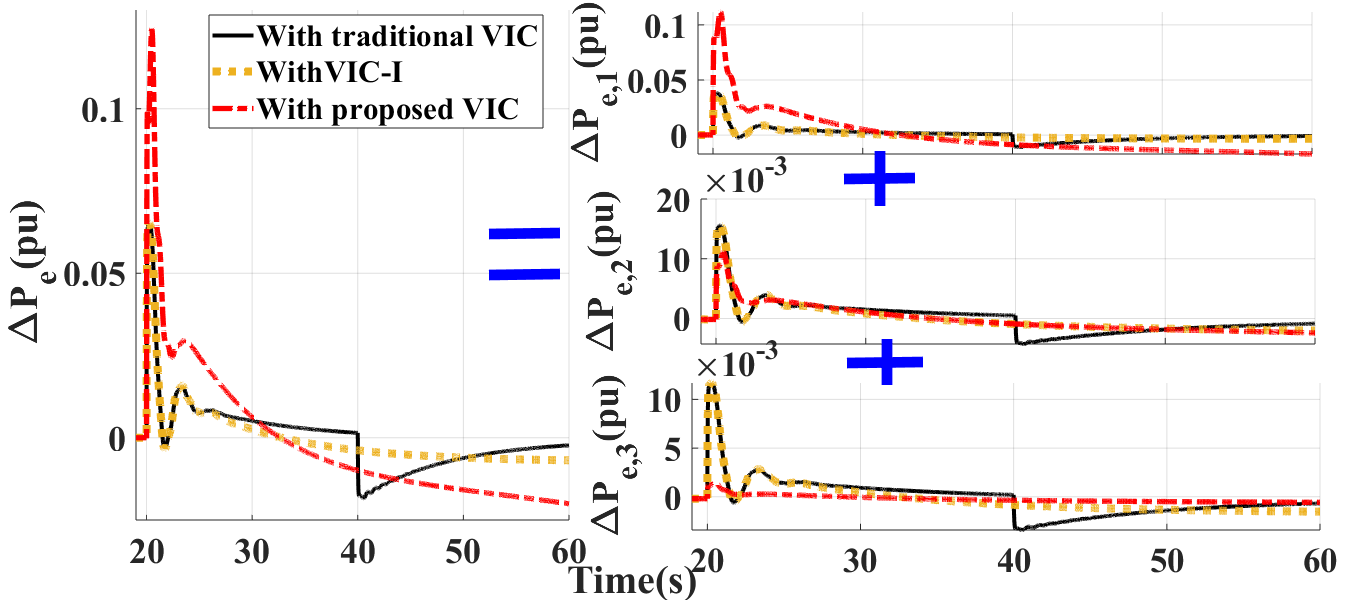}
  \caption{Active power increment of WTGs.}
  \label{Fig.14}
\end{figure}

To demonstrate the scalability of the proposed method for multiple WTGs, three WTGs with the same parameters as studied in Section III-A are integrated into the power grid. The wind speeds are assumed to be 10.8 m/s, 8 m/s and 7.3 m/s, respectively. The participating factors calculated by (\ref{Eq.29}) are 0.5842, 0.2362 and 0.1796. The state feedback coefficients for the linear quadratic optimal control are $k_1=2.2361$, $k_2=5.9389$, $k_3=6.7687$ and $k_4=3.8128$. Note that the wind power penetration rate is 60\% at this time. Comparison results about frequency responses, rotor speed responses and active power output increments are shown in Figs.~\ref{Fig.12}-\ref{Fig.14}, respectively.


These results lead to draw similar conclusions as for the cases discussed in the previous sections, as follows:  (i) the proposed VIC can effectively suppress low frequency drive-train torsional oscillation and achieve better frequency nadir as compared to other two methods, see Fig.~\ref{Fig.12} and Fig.~\ref{Fig.13}; and (ii) for WTG1 with higher wind speed, the released rotor kinetic energy with the proposed VIC is larger than the conventional VIC and VIC-I in the initial duration of frequency support. By contrast, less reductions in rotor speeds of WTG2 and WTG3 with low wind speeds are observed, which are expected, see Fig.~\ref{Fig.13}.  As a result, WTG1 results in larger active power output increment while those of WTG2 and WTG3 are smaller. However, the proposed VIC can still provide the most total power for frequency support as reported by Fig.~\ref{Fig.14}, to improve the frequency nadir. This shows how the coordination of multiple WTGs can be effectively done by the proposed VIC.

\section{Conclusions and Future work}
This paper develops a OHFT-based nonlinear virtual inertia control for WTG. It is able to provide primary frequency support of the power grid and overcome several limitations of existing VIC methods. In particular, (i) it suppresses the low frequency drive-train torsional oscillation of WTG, leading to enhancement of the service life of WTG; (ii) it can adpatively adjust the amount of released kinetic energy of WTG rotor according to the operating point; (iii) after frequency support, it achieves fast and smooth recovery of rotor speed to the original MPP point before the disturbance; and (iv) it can effectively coordinate multiple WTGs. 

Future work will focus on extending the OHFT-based control for frequency support using multiple distributed energy resources, such as energy storage, PVs, etc.

\end{document}